\documentclass{emulateapj}
\usepackage{graphicx}
\usepackage{epsfig}
\usepackage{natbib}
\usepackage{multirow}
\newcommand{\lya}{Ly$\alpha$}

\newcommand{\cii}{C\thinspace{\sc ii}}
\newcommand{\oiii}{[O\thinspace{\sc iii}]}
\newcommand{\oii}{[O\thinspace{\sc ii}]}
\newcommand{\sitwo}{Si\thinspace{\sc ii}}
\newcommand{\nii}{[N\thinspace{\sc ii}]}
\newcommand{\eg}{{\rm e.g.,\thinspace}}
\newcommand{\peaA}{J081552.00+215623.6}  
\newcommand{\peaB}{J121903.98+152608.5} 
\newcommand{\peaD}{J145735.13+223201.8} 
\newcommand{\peaC}{J030321.41-075923.2} 
\newcommand{\peaAra}{J081552.00}
\newcommand{\peaAdec}{ +215623.6}  
\newcommand{\peaAs}{J0816+22}
\newcommand{\peaBra}{J121903.98} 
\newcommand{\peaBdec}{ +152608.5}
\newcommand{\peaBs}{J1219+15}
\newcommand{\peaDra}{J145735.13}
\newcommand{\peaDdec}{ +223201.8} 
\newcommand{\peaDs}{J1457+22}
\newcommand{\peaCra}{J030321.41}
\newcommand{\peaCdec}{ -075923.2} 
\newcommand{\peaCs}{J0303-08}
\newcommand{\Zsol}{\hbox{\thinspace Z$_{\sun}$}}
\newcommand{\kmps}{\hbox{km\thinspace s$^{-1}\,$}}

\newcommand{\feii}{Fe\thinspace{\sc ii}}

\slugcomment{ApJL, Accepted July 3, 2014}

\begin{document}

\title{Linking \lya~and Low-Ionization Transitions at Low Optical Depth \footnote{Based on observations made with the NASA/ESA Hubble Space Telescope, obtained at the Space Telescope Science Institute, which is operated by the Association of Universities for Research in Astronomy, Inc., under NASA contract NAS 5-26555. These observations are associated with programs GO-13293 and GO-12928.}}

\author{A. E. Jaskot and
	M. S. Oey
	}	
\affil{University of Michigan, Department of Astronomy, 
        830 Dennison Building, Ann Arbor, MI 48109, USA.}

\begin{abstract}

We suggest that low optical depth in the Lyman continuum (LyC) may relate the
\lya~emission, \cii~and \sitwo~absorption, and \cii* and \sitwo* emission seen in
high-redshift galaxies. We base this analysis on {\it Hubble Space Telescope} COS spectra of
four Green Pea (GP) galaxies, which may be analogs of $z>2$ \lya\ emitters
(LAEs). In the two GPs with the strongest \lya~emission, the \lya~line
profiles show reduced signs of resonant
scattering. Instead, the \lya~profiles resemble the H$\alpha$ line
profiles of evolved star ejecta, suggesting
that the \lya~emission originates from a low 
column density and similar outflow geometry. The weak \cii~absorption
and presence of non-resonant \cii* emission in these GPs support this
interpretation and imply a low LyC optical depth along the line of
sight. In two additional GPs, weak \lya~emission and strong
\cii~absorption suggest a higher optical depth. These two GPs
differ in their \lya~profile shapes and \cii*~emission strengths,
however, indicating different inclinations of the outflows to our line
of sight. With these four GPs as examples, we explain the observed
trends linking \lya, \cii, and \cii* in stacked LAE spectra, in the
context of optical depth and geometric effects.  Specifically, in some galaxies
with strong \lya~emission, a low LyC optical depth may allow \lya~to
escape with reduced scattering. Furthermore, \cii~absorption, \cii* emission, and
\lya~profile shape can reveal the optical depth, constrain the orientation of
neutral outflows in LAEs, and identify candidate LyC emitters.  

\end{abstract}


\section{Introduction}
Lyman continuum (LyC) radiation from star-forming galaxies likely
caused the reionization of the universe \citep[\eg][]{bouwens12}, but
to date, studies have identified few galaxies that are LyC emitters
\citep[LCEs; \eg][]{leitherer95, heckman01, leitet13, nestor13}. At low
redshift, the Green Pea (GP) galaxies, starbursts characterized by
intense \oiii~$\lambda$5007 emission, have emerged as LCE candidates \citep[hereafter V14]{jaskot13, verhamme14}. The high
\oiii$\lambda5007$/\oii$\lambda3727$ ratios of the GPs imply an extreme ionization parameter. However, since optically thin nebulae should
underproduce \oii~\citep[\eg][]{giammanco05, pellegrini12}, these high ratios may also indicate a low LyC optical depth. Notably,
the $z<0.3$~GPs share several properties with \lya~emitters
(LAEs) at $z>2$, including high specific star formation rates, compact
sizes, low extinction, and elevated \oiii/\oii~ratios
\citep[\eg][]{hagen14,malhotra12,gawiser07,nakajima13}. Thus,
the GPs may be outstanding analogs of both high-redshift LyC-emitting
and \lya-emitting galaxies. 

Some LAEs are known LCEs (\eg~\citealt{iwata09,nestor13}, but see \citealt{vanzella10}),  and several
studies have proposed a connection between \lya~emission and a low LyC
optical depth (\eg~\citealt{shapley03,nakajima13}; V14). Due to
the resonant nature of \lya, most galaxies have low \lya~escape
fractions \citep[\eg][]{hayes11}. The geometry and kinematics of the
neutral interstellar medium (ISM) appear to control the \lya~escape
fraction \citep[\eg][]{giavalisco96, thuan97, kunth98},
with low neutral gas covering fractions and strong outflows enabling
\lya~escape \citep[\eg][]{shapley03, kornei10}. While many numerical
models have considered the effects of strong scattering through the ISM or
outflows on \lya~line profiles \citep[\eg][]{ahn00, zheng02,
  verhamme06, orsi12}, recent studies have focused on
\lya~radiative transfer through an optically thin medium (\citealt{behrens14}; V14). By driving outflows and ionizing the
surrounding ISM and circumgalactic medium, feedback from the intense
star formation in some galaxies, like the GPs, may create optically thin conditions that
enable LyC and \lya~escape.  

In this {\it Letter}, we suggest that LyC optical depth and outflow geometry link the \lya~profiles, \cii~absorption, and \cii* emission observed in high-redshift galaxies. We demonstrate this connection using {\it Hubble Space Telescope (HST)}
Cosmic Origins Spectrograph (COS) observations of four of the most
extreme GPs, galaxies that closely resemble high-redshift LAEs. We report the detection of \lya~emission in all four GPs. Three GPs show P-Cygni \lya~line profiles; the profile shapes and equivalent widths
(EWs) $>70$ \AA~in two GPs imply a relatively low optical depth. We also suggest that the enigmatic non-resonant
emission in \cii* and \sitwo* that is observed to correlate with
\lya~emission from starbursts is consistent with the geometry, optical
depth, and viewing orientation associated with the GPs' galactic
outflows. We will show that the \lya~emission strengths, \lya~profile
shapes, and weak interstellar absorption lines in two 
GPs demonstrate that some starbursts with high \oiii/\oii~may be
optically thin to the LyC. 

\section{Results}
\label{sec:results}

We present {\it HST} COS spectra of two GPs (\peaA~and \peaD) and analyze archival observations for two additional GPs (\peaB~and \peaC). \citet{jaskot13} identify these four starbursts as LCE candidates due to their high \oiii/\oii~ratios. All four GPs have similar optical spectra, metallicities ($\sim0.2$\Zsol), extinctions ($E(B-V)=0.06-0.11$), and young ages (H$\alpha$ EW$=700-1300$ \AA), but differ markedly in their UV spectra (Figures~\ref{fig_lya} and \ref{fig_cii}; Table~\ref{tablesummary}). We bin the COS spectra to $0.15-0.23$ \AA~per pixel, the resolution for emission at the GPs' NUV half-light radii (Table~\ref{tablesummary}), as measured from the COS acquisition images (\eg~James et al., in prep). Galactic \sitwo~$\lambda$1526.71 absorption lines show that the wavelength scale is accurate to within one binned pixel. At the GPs' redshifts, the COS aperture samples 6-8 kpc and fully encompasses the detected NUV emission. However, \lya~emission may originate at larger radii \citep[\eg][]{hayes13}. Based on the GPs' NUV Petrosian radii \citep{petrosian76} at the $\eta=0.2$~surface brightness level and the average ratio of \lya~radii to FUV radii in \citet{hayes13}, we expect the COS aperture to capture all of the \lya~emission. Even adopting the ratio for the most extended \lya~halo in \citet{hayes13}, the COS aperture should probe $60-100\%$~of the \lya~Petrosian radius and recover most of the \lya~flux. We use the redshifts given in the Sloan Digital Sky Survey (SDSS) Data Release 9 (\citealt{ahn12}), which are based on the GPs' optical emission lines. We adopt a cosmology with $H_0=70$ \kmps~Mpc$^{-1}$, $\Omega_m=0.3$, and $\Lambda_0=0.7$.

\subsection{\lya~Emission}

\begin{figure*}
\epsscale{1}
\plotone{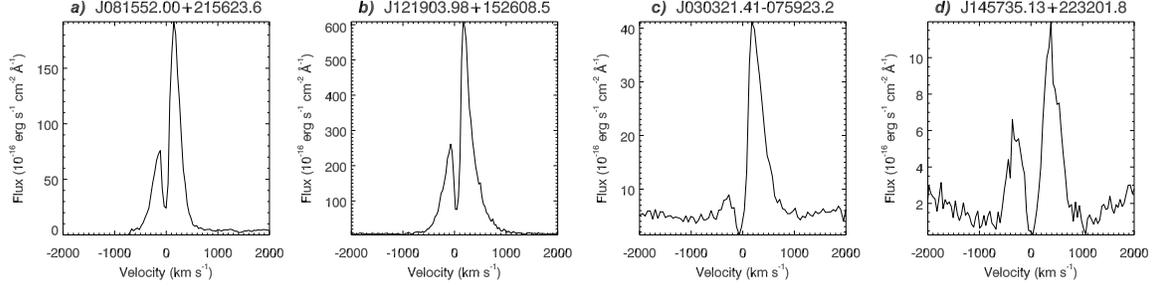}
\caption{\lya~emission in the four GPs.}
\label{fig_lya}
\end{figure*}

\begin{figure*}
\epsscale{1}
\plotone{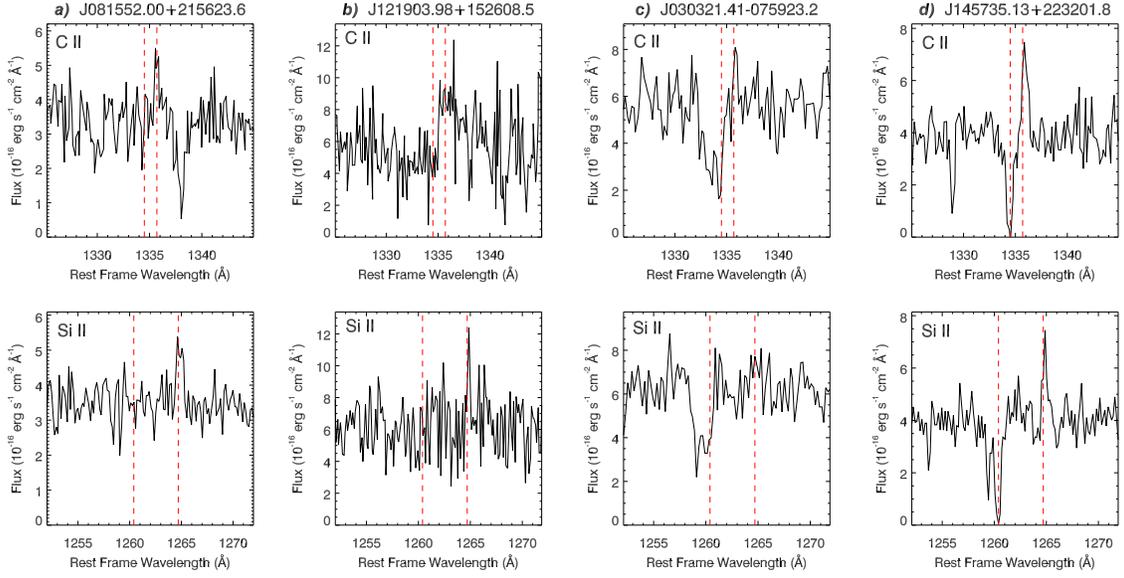}
\caption{Top panels: \cii~$\lambda$1334.5 absorption and \cii* $\lambda$1335.7 emission in the GPs. Bottom panels: \sitwo~$\lambda$1260.4 absorption and \sitwo* $\lambda$1264.7 emission. Red dashed lines indicate the expected positions of these transitions from the SDSS systemic redshifts.}
\label{fig_cii}
\end{figure*}

In the conventional scenario for \lya\ emission, \lya~scatters
many times before escaping, which significantly alters and broadens the
original line profile. However, the \lya\ profiles in the two
GPs with the highest \oiii/\oii~ratios, \peaAs~and \peaBs, do not show
the predicted effects of radiative transfer at high column density. The \lya~profiles in these two
GPs resemble a Gaussian with P Cygni-like absorption superimposed at
the systemic velocity; both galaxies have a small separation
($<$300 \kmps) of the \lya~peaks,
(Figure~\ref{fig_lya}), as expected for a neutral column density $N_{\rm HI}<10^{18}$ cm$^{-2}$ and optical depth at the Lyman edge $\tau\lesssim6$ (V14).  Indeed, V14 note that a density-bounded scenario could explain the \lya~profile of \peaBs.  

In contrast, the \lya~profiles in the other two GPs, \peaCs~and \peaDs, are consistent with higher \lya~optical depths. \peaCs\ exhibits a classic P-Cygni profile, with deeper, blue-shifted absorption, a weaker blue peak, and a greater separation of the emission peaks (Figure~\ref{fig_lya}c) than \peaAs~and \peaBs, implying a higher optical depth. This GP appears slightly more extended than the others and consists of multiple UV-emitting knots; thus, the COS aperture may miss some of the scattered emission. \peaDs~appears to have the highest line-of-sight column density, as indicated by the $\sim$750 \kmps~velocity separation of its emission peaks and the broad absorption trough extending to either side of the \lya~emission (Figure~\ref{fig_lya}d). The weak, double-peaked profile of \peaDs~resembles models of \lya~emission from highly inclined galaxies \citep{verhamme12}.

The P-Cygni \lya~profiles of \peaAs, \peaBs, and \peaCs~are strikingly similar to H$\alpha$~emission line profiles observed from stellar sources, such as symbiotic binaries (\eg~Fig. 4 of \citealt{burmeister09}) and luminous blue variables (LBVs; \eg~Fig. 4 of \citealt{weis03}). Double-peaked H$\alpha$~emission lines with weakly blue-shifted absorption are particularly common in symbiotic binaries \citep[\eg][]{quiroga02,burmeister09}, which consist of a hot compact star interacting with the wind of a red giant. The H$\alpha$~emission forms in the gas ionized by the hot star, while self-absorption from neutral gas in the red giant's wind may cause the central absorption dip \citep[\eg][]{ivison94,quiroga02}. Similar Balmer line profiles arise in LBVs \citep[\eg][]{nota97,weis03} and proto-planetary nebulae \citep[\eg][]{balick89,sanchez08}. In each of these cases, photons from an expanding ionized region or bipolar outflow encounter cooler surrounding material, often ejected by the central object. 

The narrow line profiles of \peaAs~and \peaBs~suggest a lower-than-average
\lya~optical depth and possible LyC escape. In addition, the
\lya~profiles' resemblance to the Balmer line profiles of
stellar sources indicates that they may originate from a similar
geometry, with a thin layer of neutral hydrogen obscuring a compact
ionizing source. As a resonant transition, \lya~photons generally
experience numerous scatterings. In contrast, the similarity of the
GPs' \lya~to emission profiles from a {\it non-resonant}~transition
argues that \lya~is escaping relatively unimpeded from these objects, with a reduced number of scatterings. 

\begin{figure*}
\epsscale{1}
\plotone{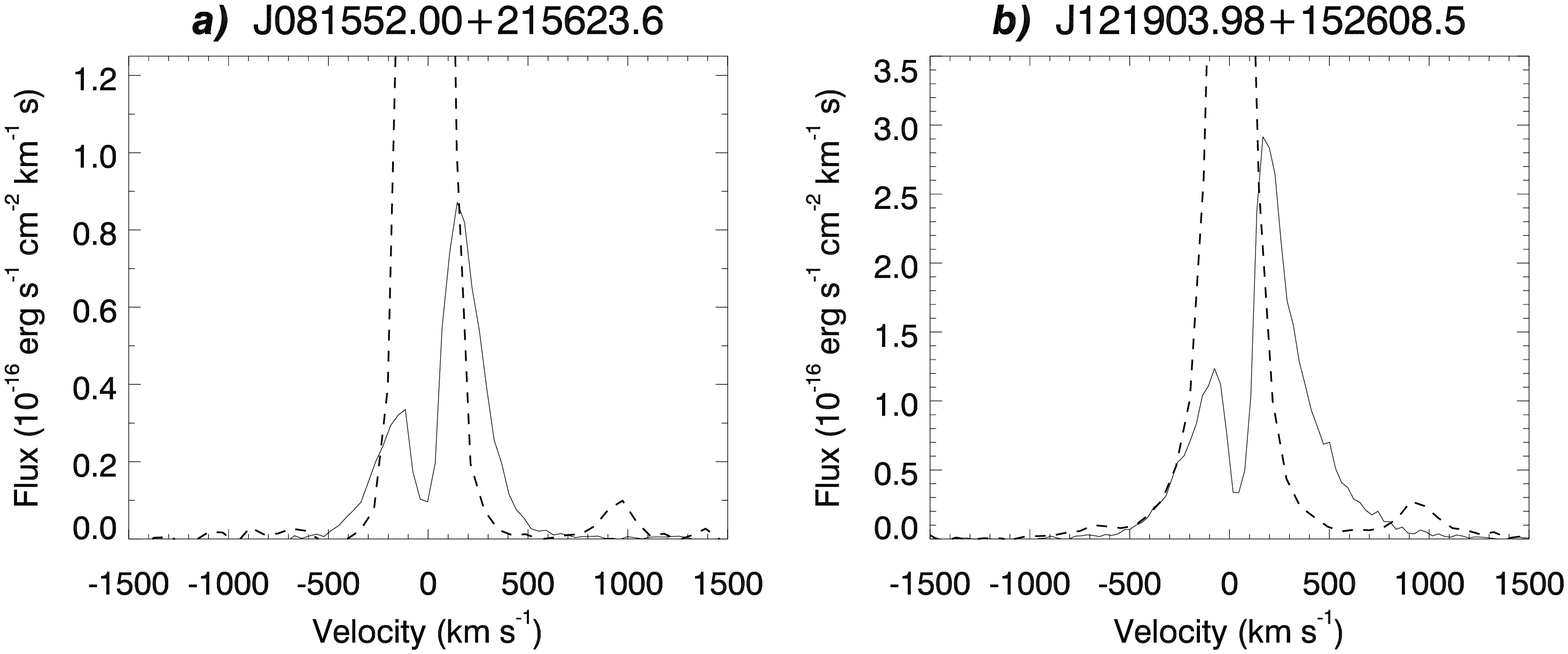}
\caption{Solid lines show the \lya~emission for \peaAs\ and \peaBs. Dashed lines show the H$\alpha$~profiles from the SDSS spectra, scaled by a factor of 8.7 to approximate the intrinsic \lya~profiles. The \nii~lines appear as bumps near -700 and +1000 \kmps. }
\label{lya_ha}
\end{figure*}

\begin{figure*}
\epsscale{1}
\plotone{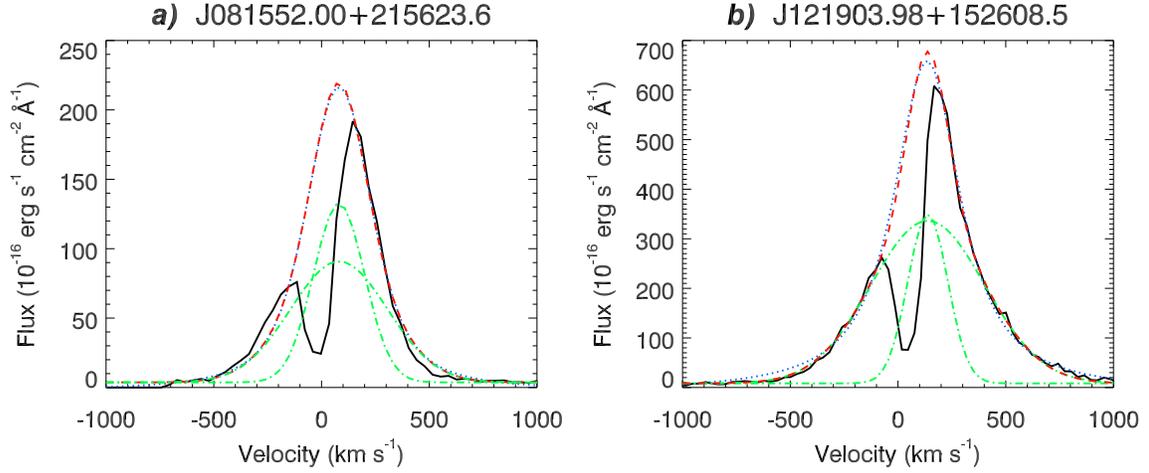}
\caption{Solid lines show observed \lya~emission. Blue dotted lines indicate a Voigt fit, and red dashed lines indicate fits based on the sum of the two Gaussian components shown with green dash-dotted lines.}
\label{lya_fits}
\end{figure*}

To illustrate this point, we estimate the intrinsic \lya\ profiles of \peaAs\ and \peaBs\ from their SDSS H$\alpha$~profiles (Figure~\ref{lya_ha}) scaled by the Case B \lya/H$\alpha$~ratio of 8.7 \citep{brocklehurst71}. The \lya~emission wings extend beyond the estimated intrinsic profiles, suggesting scattered emission. Similarly, scattering likely causes the broad H$\alpha$~wings in symbiotic star spectra \citep[\eg][]{lee00}. The \lya~profile still preserves the signature of the intrinsic emission, however, and the \lya~peaks are located near or within the intrinsic profile. In Figure~\ref{lya_fits}, we neglect the central absorption and fit the \lya~emission of \peaAs\ and \peaBs\ with a two-component sum of Gaussian functions or with a Voigt profile. Based on the adjusted $R^2$ goodness-of-fit statistic, these models are preferable to a single Gaussian fit, although the improvement is marginal for \peaAs. The narrow components of the double-Gaussian fits have FWHM values of 280 \kmps\ for \peaAs\ and 220 \kmps\ for \peaBs\, and the Gaussian components of the Voigt fits have FWHM=230 \kmps\ and 100 \kmps\, respectively. These velocities are similar to the FWHM of the H$\alpha$~profiles: 180 \kmps\ for \peaAs\ and 150 \kmps\ for \peaBs\, although these widths are upper limits given the $\sim$150 \kmps\ SDSS spectral resolution. The narrow peak separations and presence of narrow emission components suggest that \lya\ photons do not need to scatter to high velocities in order to escape. In addition, the strength of the blue peak indicates that this escape is due to a low column density rather than a high velocity outflow (V14).

A low line-of-sight optical depth is consistent with the high strength of the
\lya~emission in \peaAs~and \peaBs. These two GPs have rest-frame EWs
of $\sim$70 \AA~and 150 \AA~and \lya~escape fractions of 19\%~and
37\%, with our assumed \lya/H$\alpha$~ratio of 8.7. Such high EWs are rare at low redshift
\citep[\eg][]{finkelstein09,cowie10}, but fall in the range for LAEs
at $z=3$ \citep[\eg][]{ciardullo12}. \peaCs~has weaker \lya~emission
with an EW$\approx$10 \AA~and \lya~escape fraction of 2\%, consistent with its higher optical depth. As expected, the \lya~emission of \peaDs~is the weakest, with an escape fraction of $<1\%$.

\subsection{Interstellar Absorption and Emission Lines}
\label{sec:cii}
In stacked spectra of $z>2$ LAEs, the strength of non-resonant
emission lines such as \sitwo* and \cii* appears to correlate with
\lya~strength \citep[\eg][]{shapley03,berry12}.  The origin of
these lines is debated \citep[\eg][]{shapley03,erb10,berry12},
but here, we suggest that they probe the neutral ISM optical depth and
geometry. In particular, for the GPs, the behavior of this emission supports the
optical depths suggested by the \lya~profiles. 

The non-resonant \cii*~$\lambda$1335.7 and \sitwo*~$\lambda$1264.7 emission lines form when an excited electron decays to the first fine-structure level above the ground state. These lines share the same upper level as the \cii~$\lambda$1334.5 and \sitwo~$\lambda$1260.4 resonant transitions, which appear in absorption in stacked LAE spectra \citep[\eg][]{shapley03}. Since these absorption lines arise in the neutral ISM, their optical depths are related to the LyC optical depth and the ISM metallicity \citep[\eg][]{heckman01}, and they offer an indirect diagnostic of the line-of-sight (LOS) optical depth.

\peaAs~and \peaBs~show no detectable \cii~or \sitwo~absorption, consistent with a low optical depth (Figure~\ref{fig_cii}). As seen in spectra of high-redshift LAEs, both GPs show \cii*~and possible \sitwo*~emission. In contrast, \peaCs~and \peaDs~have clear absorption in \cii~and \sitwo. The absorption in \peaCs~is broad and blueshifted, implying the presence of an outflow, while the narrow emission and absorption in \peaDs~originate from a relatively static absorbing column. The width of the absorption lines in \peaCs~is not due to a lower spectral resolution, since the \cii~and \sitwo~absorption lines are four times wider than the observed Milky Way \sitwo~absorption. \peaCs~also has the lowest metallicity of the four galaxies \citep{izotov11}, which implies that the relative absorption strengths in the GPs do not result solely from metallicity effects. \peaCs~and \peaDs~do differ in their \cii* emission, however. While \peaCs~does not have clear \cii* emission, \peaDs~exhibits the strongest \cii* emission of the four GPs. We summarize the spectral line strengths of the GPs in Table~\ref{tablesummary}. In the following discussion, we refer to \cii, but the results are equally applicable to \sitwo.

Models for \feii~transitions developed by \citet{prochaska11} suggest an interpretation of the \cii*~emission that is entirely consistent with the geometries and optical depths implied by the GPs' \lya~emission and ISM absorption lines. Like \feii~$\lambda$2586 and \feii*~$\lambda$2612, the non-resonant \cii* and \sitwo* transitions have a slightly higher transition probability than the corresponding \cii~and \sitwo~resonant transitions. In a cool gas outflow, each \cii~absorption should be balanced by \cii~or \cii* emission into a random direction. We will therefore observe the wind component along our LOS in \cii~absorption, with weaker \cii* emission. However, neutral gas outside of the LOS to the central starburst will also absorb far-UV photons and emit some of the resulting \cii~and \cii* photons into our LOS. This emission from other parts of the wind should partially fill in the \cii~absorption and strengthen the \cii*~emission. {\it Therefore, the \cii~absorption and fluorescent \cii* emission probe the outflow geometry and optical depth}.

The lack of \cii~absorption and the presence of \cii*~emission in
\peaAs~and \peaBs~resemble the \citet{prochaska11} model for an
unobscured UV emission source. In this case, neutral gas located {\it
  outside} the LOS to the starburst absorbs UV radiation and emits
\cii*~photons {\it into} our LOS. \cii~$\lambda$1334.5 emission should be weaker than \cii* $\lambda$1335.7 due to its lower transition probability and higher chance of absorption. As discussed previously, the \lya~profiles of \peaAs~and \peaBs~also suggest minimal absorption along the LOS, as expected for a mostly unobscured source. 

\peaCs~shows broad, blue-shifted \cii~absorption, but no \cii*~emission, similar to expectations for a collimated neutral outflow aligned with the LOS \citep{prochaska11}. Neutral, outflowing gas in front of the UV emission source will cause strong \cii~absorption. This gas will also produce \cii*~emission, but as the gas will emit isotropically, most of the emission will be directed away from our LOS. The weakness of the \cii*~emission in \peaCs~would then imply a lack of neutral material outside of the outflow. A high dust optical depth could also suppress the \cii*~emission \citep{prochaska11}, but the Balmer decrement implies a low extinction in \peaCs\ \citep{jaskot13}. Alternatively, given the larger spatial extent of \peaCs, the \cii*-emitting material may be located at large radii. The COS aperture might not encompass this emission, or the lower resolution at large radii may weaken its detectability. Imaging observations may determine whether neutral gas exists outside our LOS. Regardless, the \cii~absorption, deep, blue-shifted \lya~absorption trough, and weak \lya~emission in \peaCs~are consistent with an optically thick, neutral outflow along the LOS.

While the strongest LAEs (\peaAs~and \peaBs) show \cii* emission, the strongest non-resonant emission lines appear in the weakest LAE, \peaDs. This \cii* detection is unexpected, given the trends with \cii* and \lya~strength in stacked spectra, and demonstrates that stacked spectra may not match the real spectra of individual objects. A dense ISM component along the LOS will produce strong \cii~absorption and \cii* emission near the systemic velocity \citep{prochaska11}, as seen in \peaDs. This scenario fits with our earlier conjecture that \peaDs~is highly inclined to our LOS. The high column density of the absorbing material results in a high optical depth to \cii~photons, suppressing their emission, while producing a large supply of \cii* photons that escape. This deep column of neutral gas naturally explains the broad \lya~absorption, while the weak, superimposed \lya~emission may escape from scattering in a bipolar outflow.


\section{Discussion}
\label{sec:discussion}
The geometry of the neutral ISM and its line-of-sight optical depth lead to a close connection between \lya~emission, \cii~and \sitwo~absorption, and \cii* and \sitwo* emission. Previous studies of stacked spectra of LAEs have found that higher \lya~EWs correlate with weaker interstellar absorption lines and stronger \sitwo*~emission \citep[\eg][]{shapley03,berry12}. The weak absorption lines may indicate that stronger LAEs have lower LyC optical depths \citep{shapley03}. In our spectra of individual GPs, we likewise find that the two GPs with the strongest \lya~emission may be optically thin to the LyC. Since most of their neutral gas is not located along the LOS to the central starburst, we observe fluorescent \cii* and \sitwo* emission with no accompanying \cii~and \sitwo~absorption. This geometry may account for the similar \sitwo* emission observed in stacked LAE spectra. The two GPs with weaker \lya~emission do appear to have significant neutral gas in front of the starburst, but they differ in their inclinations and geometry. As a result, while both galaxies have \cii~and \sitwo~absorption, only the more inclined GP shows strong \cii* and \sitwo* emission. \peaCs\ either lacks neutral material to the side, has high extinction, or has neutral material only at large radii. A highly extended geometry would also reduce the \lya~emission in \peaCs. If strong \cii~absorption is ubiquitous in weak LAEs, but \cii* emission depends on geometry, dust extinction, and inclination, \cii~absorption may appear in a stacked spectrum while the \cii* emission is weakened. On the other hand, the wavelength of \cii~absorption depends on outflow velocity, whereas \cii* emission should occur near the systemic velocity.

The \lya~emission and \cii~absorption in the GPs supports the hypothesis that some
galaxies with high \oiii/\oii~may be LCEs. Two of the four GPs have
weak \cii~absorption, strong \lya~emission, and \lya~profiles that
resemble Balmer emission from circumstellar ejecta. These properties
are consistent with a low LOS optical depth. The
presence of fluorescent \cii* emission suggests that these GPs do have
excited neutral gas outside the LOS.
Therefore, LyC emission may escape
anisotropically, and identifying optically thin galaxies will depend
on viewing orientation \citep[\eg][]{zastrow11,nestor11}. \peaDs,
which appears highly inclined and optically thick along the LOS, may be optically thin in other directions; emission-line
imaging may determine its transverse optical depth \citep[\eg][]{zastrow11}.

\lya~emission may be an effective diagnostic of LyC optical depth (e.g., V14). A low LyC optical depth should facilitate \lya~escape, resulting in higher \lya~EWs and \lya~line profiles that are less affected by scattering. Five of the eight $z\sim2-3$~LAEs studied by \citet{hashimoto13} have peak velocities within 200 \kmps~of the systemic velocity, similar to \peaAs\ and \peaBs, and suggestive of a low column density. The strong \lya~emission and indications of a relatively low optical depth in two GPs suggest that these galaxies are low-redshift analogs of high-redshift LAEs and LCEs. 

Some high-redshift starbursts, particularly strong LAEs, may have low neutral column densities.
As a result, these LAEs may show narrow \lya~profiles, indicating a reduced optical depth. \cii~absorption
and \cii* emission can probe the orientation and optical depth of neutral outflows in LAEs and, along with \lya~profile shape, can help identify candidate LCEs. 

\begin{table*}
\vspace*{-0.2in}
\begin{center}
\caption{GP UV Spectral Properties}
\label{tablesummary}
{\scriptsize
\begin{tabular}{lcccccccccc}
\hline
ID & $z$ & $\frac{[{\rm O\thinspace III}]}{[{\rm O\thinspace II}]}^a$ & $r_{50}^b$ & Binning & \lya~EW & $\Delta v_{\rm peak}^c$ & \cii~EW$^d$ & \cii* EW$^d$ & \sitwo~EW$^d$ & \sitwo* EW$^d$ \\ 
& & & (kpc) & (\AA~pix$^{-1}$) & (\AA) & (\kmps) & (\AA) & (\AA) & (\AA) & (\AA) \\
\hline
\peaAra & 0.1410 & 13.7 & 0.4 & 0.17 & $71\pm11$  & 260 & ... & 0.2 & ... & 0.5 \\
\peaAdec & & & & & & & & & & \\
\peaBra & 0.1956 & 12.4 & 0.5 & 0.15 & $149\pm30$  & 270 & ... & 1.4 & ... & 0.3 \\
\peaBdec & & & & & & & & & & \\
\peaCra & 0.1648 & 9.4 & 0.8 & 0.23 & $9\pm3$  & 440 & -1.3 & ... & -1.0 & ... \\
\peaCdec & & & & & & & & & & \\
\peaDra & 0.1487 & 9.8 & 0.5 & 0.18 & $-6\pm7^e$ & 750 & -0.9 & 0.7 & -0.9 & 0.3 \\
\peaDdec & & & & & & & & & & \\
\hline
\end{tabular}
\flushleft{
$^a$\oiii~$\lambda\lambda$5007,4959/\oii~$\lambda$3727 from \citet{jaskot13}. \\
$^b$NUV half-light radius. \\
$^c$Separation of \lya~peaks. \\
$^d$Based on Gaussian fits. \\
$^e$The surrounding absorption trough causes the negative EW. \\
}
\break
}
\end{center}
\end{table*}

\acknowledgments{We thank the anonymous reviewer for insightful comments that substantially improved this {\it Letter}. We are grateful to Claus Leitherer, Sangeeta Malhotra, John Salzer, and Anne Verhamme for comments on the manuscript and to Steve Finkelstein, Alex Hagen, and Tim Heckman for helpful discussions. AEJ acknowledges support from an NSF Graduate Research Fellowship. Support was provided by NASA through grant HST-GO-13293 from STScI, which is operated by AURA under NASA contract NAS-5-26555. SDSS-III is funded by the Sloan Foundation, the Participating Institutions, NSF, and DOE.

\end{document}